\documentclass[
 pre,
 reprint,
 amsmath,amssymb,
 aps,
 superscriptaddress,
]{revtex4-1}

\usepackage{graphicx}
\usepackage{bm}

\usepackage{xcolor}
\begin{document}

\title{Dynamics of radiating particles in current sheets with a transverse magnetic field component}

\author{A.A.~Muraviev}
\email{sashamur@gmail.com}
\affiliation{Institute of Applied Physics of the Russian Academy of Sciences, Nizhny Novgorod 603950, Russia}

\author{A.V.~Bashinov}
\affiliation{Institute of Applied Physics of the Russian Academy of Sciences, Nizhny Novgorod 603950, Russia}

\date{\today}

\begin{abstract}
Upcoming multipetawatt laser facilities are capable of inducing effects of quantum electrodynamics (QED) in laser-plasma interaction such as strong radiation reaction and QED cascades, both of which can significantly influence the properties and dynamics of laser plasma. This can result in the formation of extreme plasma structures with unprecedented TG levels of quasistatic magnetic fields, for example current sheets or pinch configurations of nanometer scale or smaller. In such structures radiation losses can play a significant role, so the influence of radiation losses onto the evolution of extreme current sheets deserves a separate and thorough investigation. In the current work we develop an analytical model and extend the quasiadiabatic approach describing individual particle motion onto 3D particle motion in the case when radiation reaction is non-negligible. Given that particle motion is determined by (quasi)invariants of non-dissipative motion, we derive how these (quasi)invariants evolve under the influence of radiation losses, quantify this influence and obtain a (quasi)invariant of dissipative motion. This allows reducing the dimensionality of the system of differential equations describing particle motion to just two instead of six. It is also discussed how the presented method can be used in a wider range of problems.

\end{abstract}


\maketitle

\section{\label{sec:intro}Introduction}

One of the distinguishing phenomena of relativistic laser-plasma interaction is the possibility of generation of record quasistatic magnetic fields. Different processes can cause this generation, for example: hole-boring \cite{Wilks}, self-focusing \cite{Kim}, discharge along specially shaped targets \cite{Korobkin,Korneev} or transferring of angular orbital momentum from light to matter \cite{Nuter}. An increase of laser intensity up to $10^{23}$~W cm$^{-2}$ and higher can significantly enhance magnetic field generation due to radiation reaction effects \cite{Liseykina}. Moreover, upcoming multipetawatt laser facilities \cite{Danson,Khazanov} can enable an additional unique way to produce giant magnetic fields of TG level \cite{MuravievJETPL}.

This promising way is the focus of our attention, and it is based on rapid creation of overcritical electron-positron current sheets in the field of counter-propagating laser pulses and step-like propagation of the front of plasma creation towards them \cite{MuravievJETPL} as a result of QED cascade development \cite{BellKirk}. The electric field of the laser pulse induces the electron-positron current in the last sheet before a newborn sheet screens it from the wave in less than half of the laser period. So, behind the leading sheet a comb of sheets is formed with a self-consistent distribution of currents and extremely strong quasistatic magnetic fields.

In order to determine evolution of such magnetic fields first it is necessary to study the evolution of the distribution function of particles and, as a basis, individual particle motion in this field. The primary component of the magnetic field is parallel to sheets and has a null-point in their central section. Such a structure is relevant to the pinch structure \cite{Benford} or the structure of sheets observed in the magnetosphere of planets \cite{Ness1965}. Therefore, particle motion in such structures is well studied in the case when the influence of photon emission on particle motion can be neglected \cite{Speiser1965,Sonnerup1971,Gratreau,Meierovich1986,Buchner1989,ZelenyiUFN2013,Frank}. However, recent studies with regard to laser fields \cite{Gonoskov2022,Blackburn,Popruzhenko} suggest that if electric (magnetic) fields exceed $10^{16}$~V~m$^{-1}$ (0.1~TG) this recoil can significantly affect both individual and collective ultrarelativstic particle dynamics and lead to counter-intuitive effects like enhancement of electron acceleration \cite{Zeld,BulanovPPR} or anomalous radiative trapping \cite{ART,PukhovPRL,Fedotov,Kirk_2016}. Similarly, extreme fields can make dynamics of sheets more complex or even qualitatively different due to the strong influence of radiative recoil on particle motion. A few papers were devoted to this problem. It was numerically investigated which instabilities can develop in cosmic \cite{Jaroschek2009,Hakobyan_2019} and laser-generated \cite{EfimenkoSR} sheets when radiation losses are sufficient. In relation to plasma pinching radiative collapse was analyzed based on non-dissipative trajectories \cite{Meierovich1986,Winterberg,MEIEROVICH1982} and numerically simulated \cite{EfimenkoPRE}. However, in order to understand microphysics of dissipative current sheets, radiative particle dynamics need to be studied in more detail.

Recently, we made the first step towards analytical accounting for radiative effects in individual motion of particles in current sheets \cite{Traj}. We derived an adiabatic invariant and obtained the solution of motion equations in the case when radiation losses are noticeable. However, our study has been carried out in a model, 1D case, when the magnetic field has a single component parallel to sheet and perpendicular to the current. In real conditions in both cosmic and laboratory magnetic structures the magnetic field usually has a more complex structure with components parallel to the current, as well as perpendicular to the sheet \cite{ZelenyiUFN2013,EfimenkoSR}. To our knowledge particle motion in such magnetic structures has not been studied with radiation losses taken into account. Our present step is to consider particle motion with noticeable radiation losses in a fixed two-component magnetic field. The primary component of the magnetic field is considered to be parallel to the sheet and perpendicular to the current and to linearly depend on the distance to the sheet, and the weak secondary component is considered to be constant and perpendicular to the sheet.

Note that the pre-existing approach of the guiding center \cite{Alfven,Bogoliubov} is inapplicable to describing particle motion in such magnetic field structures due to the small scale of inhomogeneities compared to the Larmor radius near the center of the sheet, where the main component of the magnetic field turns to zero. Particle motion in such structures is more elaborate, involving loop-like trajectories instead of Larmor helixes, and also chaotization of motion can occur \cite{Chen,Delcourt}. For analytical consideration of such motion the quasiadiabatic approach has been developed, where the role of quasiadiabatic invariants was designated to particle momentums averaged over fast oscillations induced by the main component of the magnetic field \cite{Sonnerup1971,Buchner1989,ZelenyiUFN2013}. But this approach neglects radiation losses. In our paper we extend it onto the case of an extreme current sheet with a fixed two-component magnetic field, when radiation losses are noticeable, and show that the phase space dimensionality can be reduced based on the extended approach. This allows simplifying the analysis of extreme current sheet dynamics.

The structure of this paper is as follows. In Section II we rederive some of the results of the work \cite{ZelenyiUFN2013} in the relativistic case and offer a way to deal with significantly different time scales of interdependent features of particle dynamics. In Section III we extend the quasiadiabatic approach onto the case with radiation reaction in the Landau-Lifshitz form \cite{LL}. In Section IV we discuss how some of the assumptions and restrictions we had employed can be lifted and show how this method can be applied in a wider range of problems.

\section{\label{sec:base}Base model}

In this section we consider \textit{without radiation losses} the motion of a \textit{relativistic} positron in a constant inhomogeneous magnetic field with two non-zero components
\begin{equation}
\label{eq:B}
\left\{\begin{array}{l}B_y(x)=kx
\\
B_x=const\end{array}\right.,
\end{equation}
and an according vector potential $A_z(x,y)=B_xy-kx^2/2+const$. The motion of an electron is symmetrical and can be derived by replacing $e$ with $-e$.

Particle motion in a field of such configuration has been studied before in \cite{Speiser1965,Sonnerup1971,ZelenyiUFN2013} in the non-relativistic case. It was shown that in the case when $B_x\ll{B_y}_{max}$ the motion can be considered as a superposition of two motions. The "slow" quasi-closed-loop motion near the sheet plane $x=0$ is due to the magnetic field component $B_x$ perpendicular to the sheet, this motion can be considered in the phase space $(y,p_y)$. The "fast" oscillations in the transverse direction are due to the $B_y$ component, which has a null point in the center of the sheet. This motion can be considered in the phase space $(x,p_x)$ \cite{ZelenyiUFN2013}.

We intend to show that in this problem relativity can be accounted for with rather minor adjustments. The key factor here is that, since the field is purely magnetic and in this section we are not considering radiation losses yet, the kinetic energy and the Lorentz-factor $\gamma$ of the particle remain constant, as well as the absolute values of momentum $p$ and velocity $V$, any of which can be considered the first integral of motion.

Additionally, for this system a second integral of motion can be obtained: since the vector-potential $A=(0,0,A_z(x,y))$ does not depend on $z$, it is evident that $P_z=p_z+eA_z/c=const$, where $P$ is the generalized momentum, $e>0$ is the positron charge, and $c$ is the speed of light. Therefore, $p_z=P_z-e(B_xy-kx^2/2)/c$. Using this expression, the system of equations for the motion of the particle can be written as:
\begin{equation}
\label{systembase}
\left\{\begin{array}{l}\dot{x}=p_x/m\gamma
\\
\dot{y}=p_y/m\gamma
\\
\dot{z}=p_z/m\gamma
\\
\dot{p_x}=-\frac{e}{c}kxp_z/m\gamma
\\
\dot{p_y}=\frac{e}{c}B_xp_z/m\gamma
\\
p_z=P_z-\frac{e}{c}(B_xy-\frac{kx^2}{2})
\end{array}\right.,
\end{equation}
where $m$ is the positron mass. Due to the conservation of total momentum the system of equations is similar to the non-relativistic one with the exception of the $1/\gamma$ factor.

From this system we can extract the systems of equations for "fast" and "slow" motion, respectively:
\begin{equation}
\left\{\begin{array}{l}\dot{x}=p_x/m\gamma
\\
\dot{p_x}=-\frac{e}{c}kx(P_z-\frac{e}{c}(B_xy-\frac{kx^2}{2}))/m\gamma\end{array}\right.,
\end{equation}
\begin{equation}
\label{ypy}
\left\{\begin{array}{l}\dot{y}=p_y/m\gamma
\\
\dot{p_y}=\frac{e}{c}B_x(P_z-\frac{e}{c}(B_xy-\frac{kx^2}{2}))/m\gamma\end{array}\right..
\end{equation}
As we can see, these systems are not completely separable, i.e. $\dot{p_y}$ depends on $x$, and $\dot{p_x}$ depends on $y$. However, the assumption of significantly different timescales for the "fast" and the "slow" motion allows us to separate them.

For the sake of "fast" motion $y$ and $p_y$ are "frozen" external parameters. Thus, the conservation of momentum during "fast" motion can be written as ${p_x}^2+{p_z}^2=p^2-{p_y}^2=const$\cite{ZelenyiUFN2013}, since $p_y$ does not change on the "fast" time scale. Then the parameter $\eta$ from \cite{Traj} (or $-s$ from \cite{ZelenyiUFN2013}) that determines the type of motion in the "fast" phase space $(x,p_x)$ can be written as
\begin{equation}
\label{eq:eta}
\eta(y,p_y)=(P_z-eB_xy/c)/\sqrt{p^2-{p_y}^2}
\end{equation}
\footnote{This expression can also be derived by the same logic as in \cite{ZelenyiUFN2013} for the parameter $s$ (or $-\eta$), with a remark that the Hamiltonian can only be divided into terms associated with the motion in the $(x,y,z)$ direction due to the conservation of the common Lorentz-factor $\gamma$, and that these terms no longer directly represent the kinetic energy.}. In the work \cite{Traj} $\eta$ was the second integral of motion as there was no $y$ or $p_y$. In this work, however, the current position of the particle on the "slow" $(y,p_y)$ plane corresponds to and determines $\eta$ and the form of the trajectory on the "fast" $(x,p_x)$ plane. 

For the sake of "slow" motion the terms depending on the "fast" variable $x$ can be averaged over the period of "fast" motion, and the averaged values are functions of $(y,p_y)$, which can be computed with the help of the results of \cite{Traj}. For example, as follows from \cite{Traj} Eq.2,
\begin{equation}
\label{x^2}
\overline{x^2}=\frac{2c}{ek}p\sqrt{1-{q_y}^2}(\overline{\cos{\varphi}}-\eta) ,
\end{equation}
where $q_y\equiv p_y/p$ \footnote{Both $p_y$ and $q_y$ will be used throughout the paper, but in most arguments they are interchangeable.} and $\varphi$ is the "fast" motion direction of propagation (projected onto the $xz$ plane): $\sin{\varphi}=p_x/p_{xz}$, $\cos{\varphi}=p_z/p_{xz}$, where $p_{xz}=\sqrt{p_x^2+p_z^2}$. The value $\overline{\cos{\varphi}}$ can be computed as
\begin{equation}
\overline{\cos{\varphi}}=\frac{\int_{T_x} {\cos{\varphi}} dt}{\int_{T_x} dt}=\frac{\int_0^{\varphi_{max}} (\cos{\varphi}/\dot{\varphi}) d\varphi}{\int_0^{\varphi_{max}} (1/\dot{\varphi}) d\varphi},\nonumber
\end{equation} where $T_x$ is the period of "fast" motion on the $(x,p_x)$ plane.

Since $\dot{\varphi}=\sqrt{2ek/m\gamma}\sqrt{\cos{\varphi}-\eta}$ \cite{Traj},
\begin{equation}
\overline{\cos{\varphi}}=\frac{\int_0^{\varphi_{max}} \frac{\cos{\varphi}}{\sqrt{\cos{\varphi}-\eta}} d\varphi}{\int_0^{\varphi_{max}} \frac{1}{\sqrt{\cos{\varphi}-\eta}} d\varphi} ,\nonumber
\end{equation}
which with the help of the substitution $\sin{\alpha}=\sin{(\varphi/2)}/\sqrt{(1-\eta)/2}$ can be computed as
\begin{equation}
\label{eq:cos}
\overline{\cos{\varphi}}=
\left\{\begin{array}{l}2\frac{E[(1-\eta)/2]}{K[(1-\eta)/2]}-1, \eta>-1
\\
\eta-(1+\eta)\frac{E[2/(1+\eta)]}{K[2/(1+\eta)]}, \eta<-1\end{array}\right.,
\end{equation}
where $K$,$E$ are complete elliptic integrals of the first and second kind, respectively, and $\eta$ is a function of $(y,p_y)$ as defined in (\ref{eq:eta}). The notation
\begin{equation}
\label{eq:elliptic}
\left\{\begin{array}{l}K[\kappa^2]=\int_0^{\pi/2} d\varphi/\sqrt{1-\kappa^2\sin^2\varphi}
\\
E[\kappa^2]=\int_0^{\pi/2} d\varphi\sqrt{1-\kappa^2\sin^2\varphi}\end{array}\right.
\end{equation}
is used.

\begin{figure}[b]
\includegraphics[width=\columnwidth]{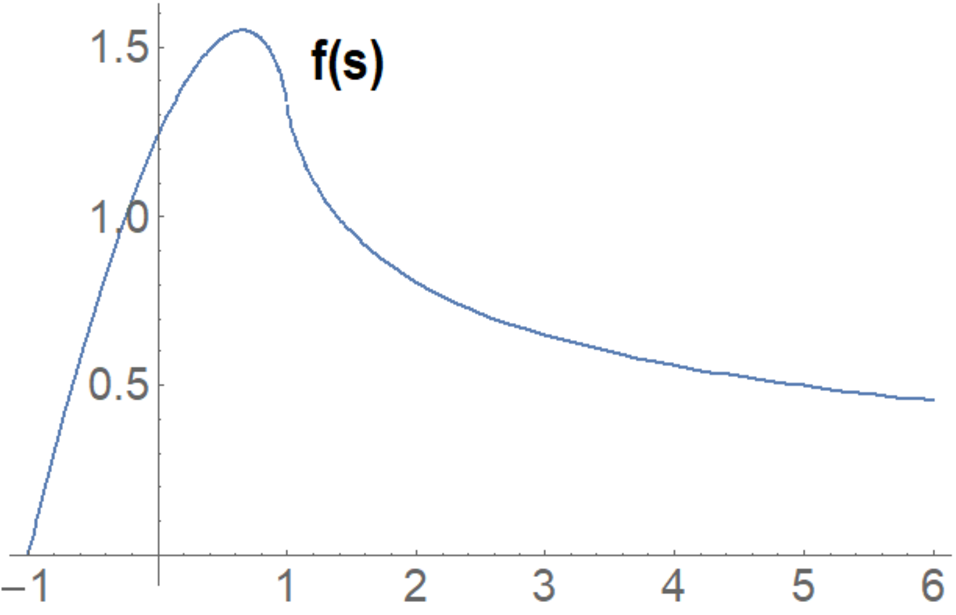}
\caption{\label{f(s)} The function f(s).}
\end{figure}

In theory equations (\ref{eq:eta}), (\ref{x^2}) and (\ref{eq:cos}) allow writing the term $x^2$ as a function of $(y,p_y)$ and thus excluding $x$ from the system of equations (\ref{ypy}), but one can see that the expressions are rather complex and solving (\ref{ypy}) directly would be problematic. Therefore, in order to study the system (\ref{ypy}) we employ the quasiadiabatic invariant \cite{ZelenyiUFN2013}:
\begin{equation}
\label{ix}
i_x(y,q_y)=\frac{1}{2\pi}(1-{q_y}^2)^{3/4}f(-\eta) ,
\end{equation}
where
\begin{equation}
\label{eq:f}
f(s)=
\left\{\begin{array}{l}\frac{2}{3}(K[\kappa^2](1-s)+2sE[\kappa^2]), -1\leq s\leq 1
\\
\frac{4}{3}\kappa(K[\kappa^{-2}](1-s)+sE[\kappa^{-2}]), s\geq 1 \end{array}\right.
\end{equation}
and $\kappa=\sqrt{(1+s)/2}$. Unlike the work \cite{ZelenyiUFN2013}, in our notation $i_x$ is dimensionless. For convenience, the function $f(s)$ is presented in Fig.~\ref{f(s)}.

\begin{figure}[b]
\includegraphics[width=\columnwidth]{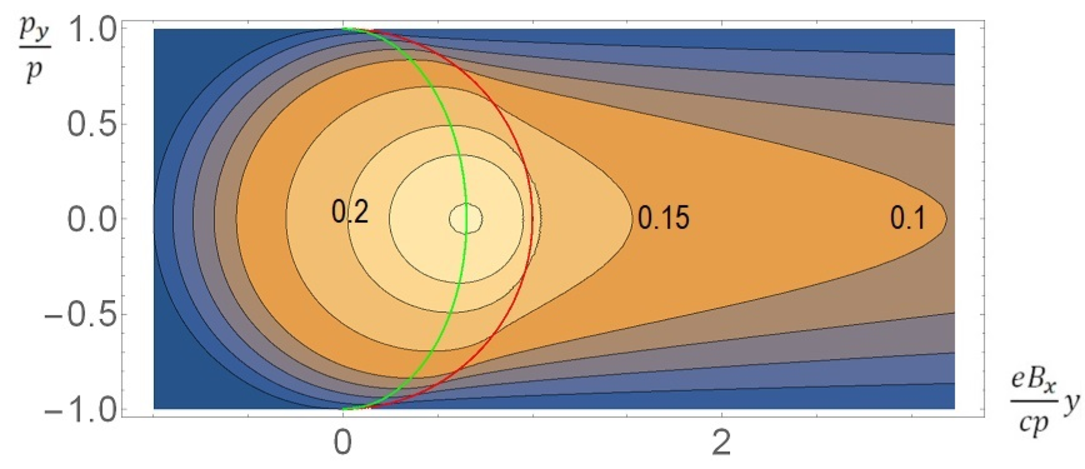}
\caption{\label{phsp} The phase space of "slow" variables $(y,q_y)$. The uncertainty curve is shown in red. The green curve represents $\dot{p_y}=0$. Trajectories are denoted with the corresponding value of $i_x$.}
\end{figure}

Each possible value of $i_x$ corresponds to a closed loop on the $(y,q_y)$ plane, the possible values are $0<i_x<i_{xmax}\equiv f_{max}/(2\pi)\approx0.246$, where $f_{max}\approx1.55$ is the maximal value of the function $f(s)$. $i_x$ is well conserved along the trajectory except in the vicinity of the \textit{uncertainty curve} (in red on Fig.~\ref{phsp}) determined by $\eta(y,q_y)=-1$, where it experiences a quasirandom jump \cite{ZelenyiUFN2013}. Since this paper is devoted to studying the influence of radiation losses on the trajectory, and not much radiation loss is accumulated in the small vicinity of the uncertainty curve where the jump occurs, the quasirandom jumps of the quasiadiabatic invariant will be disregarded throughout this paper. The green curve on Fig.~\ref{phsp} is the curve $\dot{q_y}=0$ or $\eta(y,q_y)=\eta^*$ \cite{Traj}, which will be shown to be of importance further. To the left of the green curve $\eta(y,q_y)$ takes values $\eta^*<\eta(y,q_y)\leq1$ and the "fast" trajectory is of type A or B from \cite{Traj} (meandering). Between the green and the red curves $-1<\eta(y,q_y)<\eta^*$ and the trajectory is of type C \cite{Traj}. To the right of the red curve $\eta(y,q_y)<-1$ and the trajectory is of type D \cite{Traj} (non-crossing).

An interesting feature of this system is that $\dot{z}{\sim}\dot{p_y}{\sim}p_z$. As a result, with the right choice of the origin of coordinates $z{\sim}p_y$, and the trajectory in the phase space $(y,p_y)$ is similar to the projection of the trajectory onto the $yz$ plane. In the $x$ direction the trajectory oscillates around zero if the current point on the $(y,p_y)$ plane is to the left of the red curve and around a non-zero value with a quasirandom sign \cite{ZelenyiUFN2013, Traj} if it is to the right of the red curve. Thus, the whole trajectory resembles a split torus \cite{ZelenyiUFN2013}.

\section{\label{sec:class}Implementing Radiative Recoil}

A positron moving along a curvilinear trajectory can emit photons. Based on the preceding works \cite{MuravievJETPL,Traj} we allow the magnetic field and particle energy values to be sufficiently high in order for the particles to exhibit radiative recoil. Even though a single impact experienced by the particle as a result of photon emission can be relatively weak, particle motion can be qualitatively modified as a result of a sequence of such acts. While recoil-free motion of particles would be infinite and quasiperiodic as described in Sec.\ref{sec:base}, even recoil that is insignificant over one period of particle motion may accumulate over multiple periods and have a significant effect on the motion of particles. In the work \cite{MuravievJETPL} numerical simulations predict current sheets to be formed by ultraintense laser fields. The lifetime of these current sheets was predicted to be much larger than the laser wave period, which is in turn much greater than the characteristic periods of particle trajectories, so such an accumulation may indeed take place.

In this section we consider particle motion in the field structure with two non-zero components $B_y(x)=kx$ and $B_x=const$ \textit{with radiative recoil}. In the case when particles emit photons often and they carry away a negligible part of the particle's energy, it is reasonable to consider radiation losses in the form of a continuous Landau-Lifshitz friction force \cite{LL}. We also consider only the ultrarelativistic case $p/mc\approx\gamma\gg1$, which allows us to neglect the first and second terms of the LL force \cite{LL}. In our setup this translates to:

\begin{equation}
\label{Frad}
{\vec{F}}_{rad}=-\frac{2e^4\vec{V}}{3m^2c^7}{\gamma}^2[V\times B]^2
,
\end{equation}

or, after expanding the vector product,

\begin{eqnarray}
{\vec{F}}_{rad}=-\frac{2e^4\vec{V}}{3m^2c^7}{\gamma}^2\left[\right.k^2x^2(1-q_y^2)+\nonumber\\
+B_x^2(1-(1-q_y^2)\sin^2{\varphi})\nonumber-\\
\label{Frad2}-2kxB_xq_y\sqrt{1-q_y^2}\sin{\varphi}\left.\right]
.
\end{eqnarray}

In Sec.~\ref{sec:base} $\gamma$ was constant as there was no recoil/friction. Accounting for radiative friction forces us to treat $\gamma$ as another parameter depending on time. Like in \cite{Traj}, we will use the variable $\mu=1/\gamma$ as the main variable characterizing the current energy. Taking into account that particle energy strictly decreases due to radiative friction given by the LL force in Eq.~(\ref{Frad2}), the equation for $\dot{\mu}=\frac{d\mu}{dt}$ can be written as:

\begin{eqnarray}
\dot{\mu}=\frac{2e^4}{3m^3c^5}\left[\right.k^2x^2(1-q_y^2)+\nonumber\\
+B_x^2(1-(1-q_y^2)\sin^2{\varphi})\nonumber-\\
\label{dotmu}-2kxB_xq_y\sqrt{1-q_y^2}\sin{\varphi}\left.\right]
.
\end{eqnarray}

Thus, the main parameters characterizing the state of the particle are $\mu$ (determines the particle's energy) and $i_x$ (determines what trajectory the particle is on in the $(y,q_y)$ phase space). $(y,q_y)$ determine at which point of that trajectory that particle is at the moment, and the shape of the potential the particle oscillates in, as well as the particle's energy, in the $(x,p_x)$ phase space. It is important that $(y,q_y)$ are not independent considering a given $i_x$. And finally, $x$ or $\varphi$ determines at which point of that "fast" potential the particle is at the moment.

Our primary objective is to study the influence of \textit{weak} radiative friction, so we will assume that

\begin{equation}
\label{RRweak}
T_x\ll{}T_y\ll{}\frac{\mu}{\dot{\mu}},\frac{i_x}{\dot{i_x}}
,
\end{equation}
where $T_y$ is the period of motion on the $(y,q_y)$ plane.

The positron's motion at any given moment of time can be approximated by the solution for the case without radiative friction, and we treat $\mu$ and $i_x$ as slowly varying parameters. In order to account for radiative friction, we must quantify its influence on these parameters over the half-period of motion in the $(y,q_y)$ plane $T_y/2$ assuming constant $\mu$ and $i_x$, and find how that quantity depends on $\mu$ and $i_x$.

To quantify $\Delta\mu$ we integrate $\dot{\mu}(\mu,i_x)$ over $T_y/2$. In order to do that, we use $d\mu/dq_y=\dot{\mu}/\dot{q_y}$ and integrate over $q_y$ instead. Here, from (\ref{systembase}) and (\ref{x^2}),
\begin{equation}
\label{dotqy}
\begin{array}{l}
\dot{q_y}=\frac{eB_x\mu}{mc}((P_z-\frac{e}{c}B_xy)+\frac{ekx^2}{2c})/p=\\
=\frac{eB_x\mu}{mc}(\eta\sqrt{1-q_y^2}+\sqrt{1-q_y^2}(\overline{\cos{\varphi}}-\eta))=\\
=\frac{eB_x\mu}{mc}\sqrt{1-q_y^2}\overline{\cos{\varphi}}
\end{array}
\end{equation}
However, since $\eta$ in the expression for $\dot{\mu}$ (\ref{dotmu}) depends on both $y$ and $q_y\equiv p_y/p$ (\ref{eq:eta}), and since we are integrating over a trajectory with constant $i_x$, $\eta$ must first be written as a function of $i_x$ and $q_y$ instead. Both pairs of variables are valid ways of representing a point on the $(y,p_y)$ plane \footnote{The ambiguity due to the sign of $\dot{q_y}$ will be addressed later}. Note that radiative friction is not inflicted onto $q_y\equiv p_y/p$ since the force (\ref{Frad}) is always directed against the particle's velocity. As such, $\eta$ must be written as:

\begin{equation}
\label{eq:etaix}
\eta(i_x,q_y)=-f^{-1}(2\pi i_x(1-q_y^2)^{-3/4})
,
\end{equation}
as follows from (\ref{ix}). Here, $f^{-1}$ is the inverse function of $f$.

For simplicity here we will recall that we consider $B_x$ to be weak \cite{ZelenyiUFN2013} and neglect the second and third terms of (\ref{dotmu}). Then, $\Delta\mu$ can be written as:

\begin{equation}
\label{eq:dmu}
\Delta\mu(\mu,i_x)=\frac{4e^2k}{3mc^2B_x\mu^2}\oint_{i_x} (1-q_y^2)(1-\eta/\overline{\cos{\varphi}}(\eta)) \,dq_y
,
\end{equation}
where $\eta=\eta(i_x,q_y)$ according to (\ref{eq:etaix}) and $\overline{\cos{\varphi}}(\eta)$ is a function of $\eta$ according to (\ref{eq:cos}). The integral is performed over a closed-loop trajectory on $(y,p_y)$, on which $i_x=const$. For parts of the trajectory when $\dot{q_y}>0$ (to the left of the green curve on Fig.~\ref{phsp}) the lower branch of the two-valued function $f^{-1}$ should be used. When $\dot{q_y}<0$ (to the right of the green curve on Fig.~\ref{phsp}), its upper branch should be used.

As a result, the integral is broken up into two parts for trajectories not crossing the \textit{uncertainty curve}, those have $i_x$ in the interval $0.212\approx2/3\pi\equiv i_x^*<i_x<i_{xmax}\equiv f_{max}/2\pi\approx0.246$ \footnote{$f_{max}\approx1.547$ is the maximal value of $f(s)$. The value $i_x$ for the trajectory tangent to the \textit{uncertainty curve} can be found by setting $q_y=0$ and $\eta=-1$ in (\ref{ix})}, and into three parts for trajectories crossing the \textit{uncertainty curve} ($0<i_x<i_x^*$) because of the piecewise properties of the functions $f(s)$ and $\overline{\cos{\varphi}}(\eta)$. The limits of integration can be found by substituting $\eta=\eta^*\approx-0.65$, where $f(-\eta^*)=f_{max}$ (also see \cite{Traj}), for the green curve, and $\eta=-1$ for the red curve into (\ref{ix}) and solving for $q_y$:

\begin{equation}
\label{eq:qy}
q_y=\pm\sqrt{1-\left(\frac{2\pi i_x}{f(-\eta)}\right)^{4/3}}
.
\end{equation}

Note that in these variables the integrable part of (\ref{eq:dmu}) depends only on $i_x$, while all dependence on $\mu$ can be factored out of the integral. As a result, $\Delta\mu(\mu,i_x)$ can be written as a product of two functions with separated variables: $\Delta\mu(\mu,i_x)=F_\mu(\mu)F_{i_x}(i_x)$. In this case we can define $F_\mu(\mu)=4e^2k/3mc^2B_x\mu^2$, then
\begin{equation}
\label{eq:Fix}
F_{i_x}(i_x)=\oint_{i_x} (1-q_y^2)(1-\eta/\overline{\cos{\varphi}}(\eta)) \,dq_y
,
\end{equation}
where, again, $\eta=\eta(i_x,q_y)$ according to (\ref{eq:etaix}) and $\overline{\cos{\varphi}(\eta)}$ is a function of $\eta$ according to (\ref{eq:cos}).
We will see momentarily that the same thing applies to $\Delta i_x$, which can be written as $\Delta i_x(\mu,i_x)=G_\mu(\mu)G_{i_x}(i_x)$.

Since any change in the quasiadiabatic invariant $i_x$ is a direct consequence of the change in $\mu$ by radiative friction, $\dot{i_x}$ can be calculated as $\dot{i_x}=\frac{di_x}{d\mu}\dot{\mu}$, where $\frac{di_x}{d\mu}$ can be derived from (\ref{ix}):
\begin{equation}
\label{eq:dixdmu}
\frac{di_x}{d\mu}=-\frac{1}{2\pi}(1-q_y^2)^{3/4}f'(-\eta)\frac{\eta}{\mu}
.
\end{equation}
Accordingly, the expression for the change of $i_x$ over the half-period $T_y/2$ can be expressed as:
\begin{equation}
\label{eq:dix}
\Delta i_x=-\frac{G_\mu(\mu)}{2\pi}\oint_{i_x} (1-q_y^2)^{7/4}\eta f'(-\eta)(1-\eta/\overline{\cos{\varphi}}(\eta)) \,dq_y
,
\end{equation}
keeping in mind (\ref{eq:etaix}) and (\ref{eq:cos}). Here we chose $G_\mu(\mu)=-4e^2k/3mc^2B_x\mu^3$ or

\begin{equation}
\label{eq:fgmu}
G_\mu(\mu)=F_\mu(\mu)/\mu
,
\end{equation}
while 
\begin{equation}
\label{eq:Gix}
G_{i_x}(i_x)=-\frac{1}{2\pi}\oint_{i_x} (1-q_y^2)^{7/4}\eta f'(-\eta)(1-\eta/\overline{\cos{\varphi}}(\eta)) \,dq_y
;
\end{equation}
(\ref{eq:etaix}), (\ref{eq:cos}).

The functions $F_{i_x}(i_x)$ and $G_{i_x}(i_x)$ are quite hard to calculate due to the implicit definition of $f^{-1}$ and piecewise properties of many underlying functions, but we have now reduced them to \textit{dimensionless} functions of a single variable, which means they can be \textit{calculated once and tabulated} with any required precision.

Moreover, if we consider timescales much larger than $T_y$, the expressions $\frac{\Delta\mu}{T_y/2}$ and $\frac{\Delta i_x}{T_y/2}$ can be viewed as differential equations $\dot{\mu}$ and $\dot{i_x}$, respectively:

\begin{equation}
\label{muixdiff}
\left\{\begin{array}{l}\dot{\mu}(\mu,i_x)=\frac{F_\mu(\mu)F_{i_x}(i_x)}{T_y/2}
\\
\dot{i_x}(\mu,i_x)=\frac{G_\mu(\mu)G_{i_x}(i_x)}{T_y/2} \end{array}\right.
\end{equation}

But in that case there exists an invariant $L(\mu,i_x)=L_\mu(\mu)L_{i_x}(i_x)$. From $\frac{d}{dt}L(\mu,i_x)=0$ follows $L_\mu'(\mu)L_{i_x}(i_x)F_\mu(\mu)F_{i_x}(i_x)+L_\mu(\mu)L_{i_x}'(i_x)F_\mu(\mu)F_{i_x}(i_x)=0$ or:
\begin{equation}
\label{dL=0}
\frac{L_\mu'(\mu)}{L_\mu(\mu)}\frac{F_\mu(\mu)}{G_\mu(\mu)}=-\frac{L_{i_x}'(i_x)}{L_{i_x}(i_x)}\frac{G_{i_x}(i_x)}{F_{i_x}(i_x)}=const
,
\end{equation}
since the left side does not depend on $i_x$ and the right side does not depend on $\mu$.

In that case due to Eq.~\ref{eq:fgmu} it is easily found that
\begin{equation}
L_\mu=exp\left(\int \frac{G_\mu(\mu)}{F_\mu(\mu)} d\mu\right)=\mu
.
\end{equation}

\begin{equation}
\label{eq:L1L2}
L_{i_x}=exp\left(-\int \frac{F_{i_x}(i_x)}{G_{i_x}(i_x)} di_x\right)
.
\end{equation}

We will use Eq.~\ref{eq:L1L2} to find $L_{i_x}$ for all $i_x$ in the following subsection. But for trajectories with $i_x\ll i_{xmax}$ it can be estimated independently of Eq.~\ref{eq:L1L2} that $L_{i_x}=i_x^2$. Indeed, since $\dot{q_y}\sim\dot{z}$, and $\overline{\dot{z}}$ is very small on the near-Larmor trajectories with $-\eta\gg1$ (to the right of the red curve on Fig.~\ref{phsp}) \cite{Traj,Sonnerup1971} as compared to meandering trajectories (to the left of the red curve), and the particle completes a closed loop on the $(y,p_y)$ plane, the particle must spend most of its time with $\dot{p_y}<0$ (to the right of the green curve) rather than with $\dot{p_y}<0$ (to the left of the green curve). Additionally, its average $x^2$ is also larger \cite{Traj,Sonnerup1971,ZelenyiUFN2013}, and therefore so is its exposure to radiative friction. Therefore, for a rough estimate the influence of weaker radiative friction over shorter periods of time to the left of the red curve can be neglected, and only the part of the trajectory to the right of the red curve can be considered. Finally, remembering that $i_x\sim\oint p_x dx/p^{3/2}$ \cite{ZelenyiUFN2013}, the Larmor radius in the ultrarelativistic case $R_L\sim\gamma$ and $p\sim\gamma$, it can be concluded that $i_x\sim\gamma^{1/2}=\mu^{-1/2}$, thus $i_x^2\mu\approx const$.

\subsection{\label{sec:muix}The functions $\Delta\mu(i_x)$, $\Delta i_x(i_x)$ and $L(i_x)$}

At this point the $\mu$-dependent parts of the discussed functions have been identified and only $i_x$-dependent parts remain, so we will refer to $F_{i_x}$ as $\mu$, to $G_{i_x}$ as $i_x$, and to $L_{i_x}$ as $L$ in order to underline their physical meaning. Although exact analytical expressions for these functions were obtained (Eqs.(\ref{eq:Fix},\ref{eq:Gix},\ref{eq:L1L2}) with substitutions (\ref{eq:cos},\ref{eq:elliptic},\ref{eq:f},\ref{eq:etaix},\ref{eq:qy})), the elaborate nature of these expressions gives little idea of the actual form of the functions. As such, in this subsection we present the results of numerical calculations of the specified dimensionless functions, as well as discuss fits for these functions in different regions, and discuss the inferences.

\subsubsection{\label{sec:dmu} $\Delta\mu(i_x)$}

Using asymptotic expressions for the underlying functions in the case $i_x\ll i_{xmax}$ and $-\eta\gg1$ a dependence $\Delta\mu\sim i_x^{-4}$ can be derived, so we expect to see $\Delta\mu i_x^4\approx const$ near $i_x=0$.

\begin{figure*}
\includegraphics[width=2\columnwidth]{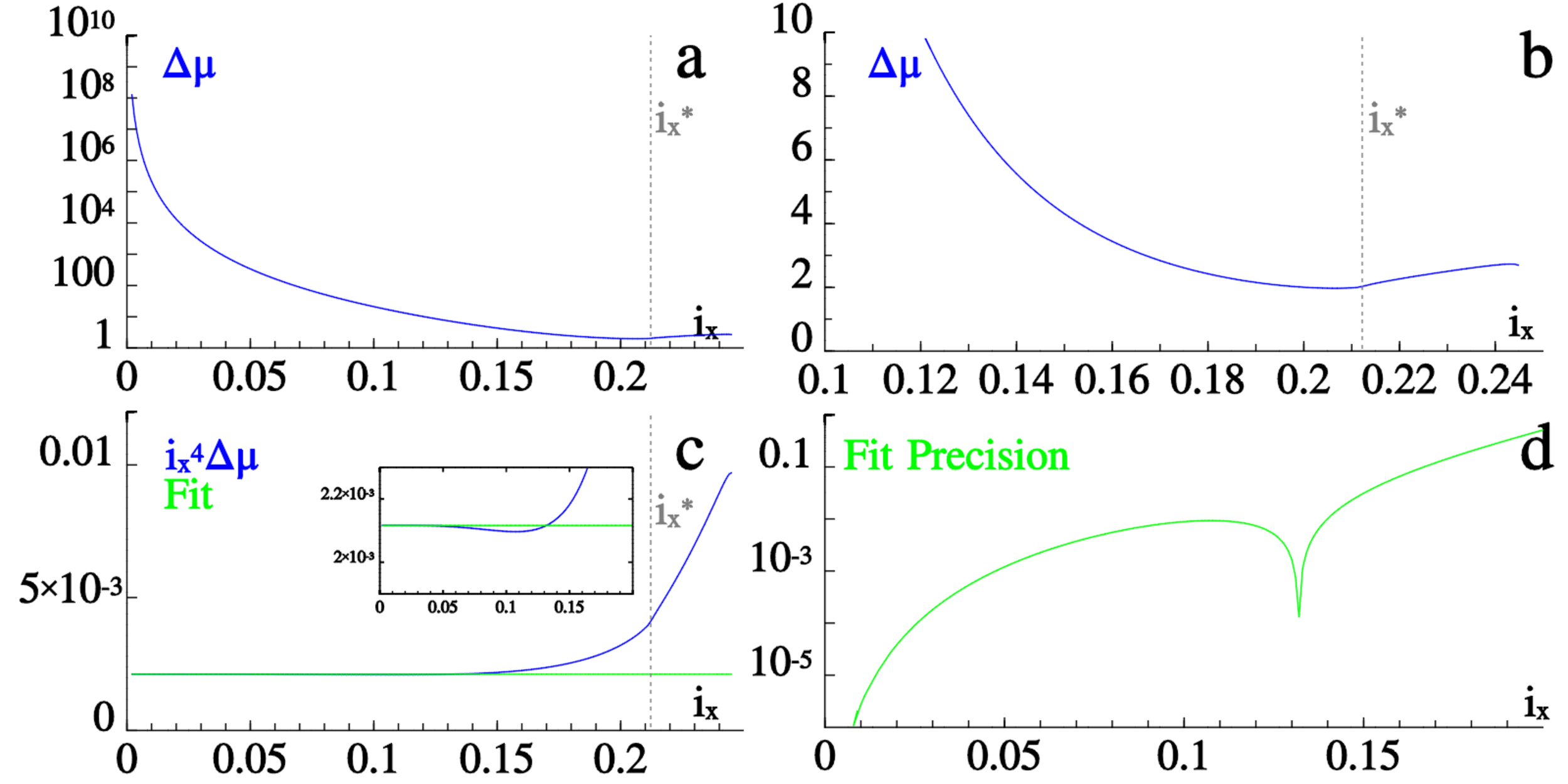}
\caption{\label{Fig:dmu} (a) $\Delta\mu(i_x)$ on a logarithmic scale. (b) The lower part of $\Delta\mu(i_x)$ ($i_x>0.1$) on a linear scale. (c) $\Delta\mu i_x^4(i_x)$ (blue) and its fit (green), in this case a constant. (d) Accuracy of the fit on a logarithmic scale. $i_x=i_x^*$ is shown in gray.}
\end{figure*}

The graph of $\Delta\mu(i_x)$ is shown in Fig.~(\ref{Fig:dmu}). It is indeed evident that $\Delta\mu i_x^4=const$ is a good approximation for $i_x\ll i_{xmax}$ and it holds a better than 1\% precision for $i_x<0.15$. Note that the function has a breaking point near $i_x=i_x^*\approx0.212$, which corresponds to the trajectory on Fig.\ref{phsp} tangent to the red curve.

\subsubsection{\label{sec:dix} $\Delta i_x(i_x)$}

Using the approximate invariant $L\approx\mu i_x^2$, or using asymptotic expressions for the underlying functions in the case $i_x\ll i_{xmax}$ and $-\eta\gg1$, a dependence $\Delta\mu\sim i_x^{-3}$ can be derived, so we expect to see $i_x^3\Delta i_x\approx const$ near $i_x=0$. Note that $\Delta i_x$ is negative, so the graph depicts the value $-\Delta i_x$.

\begin{figure*}
\includegraphics[width=2\columnwidth]{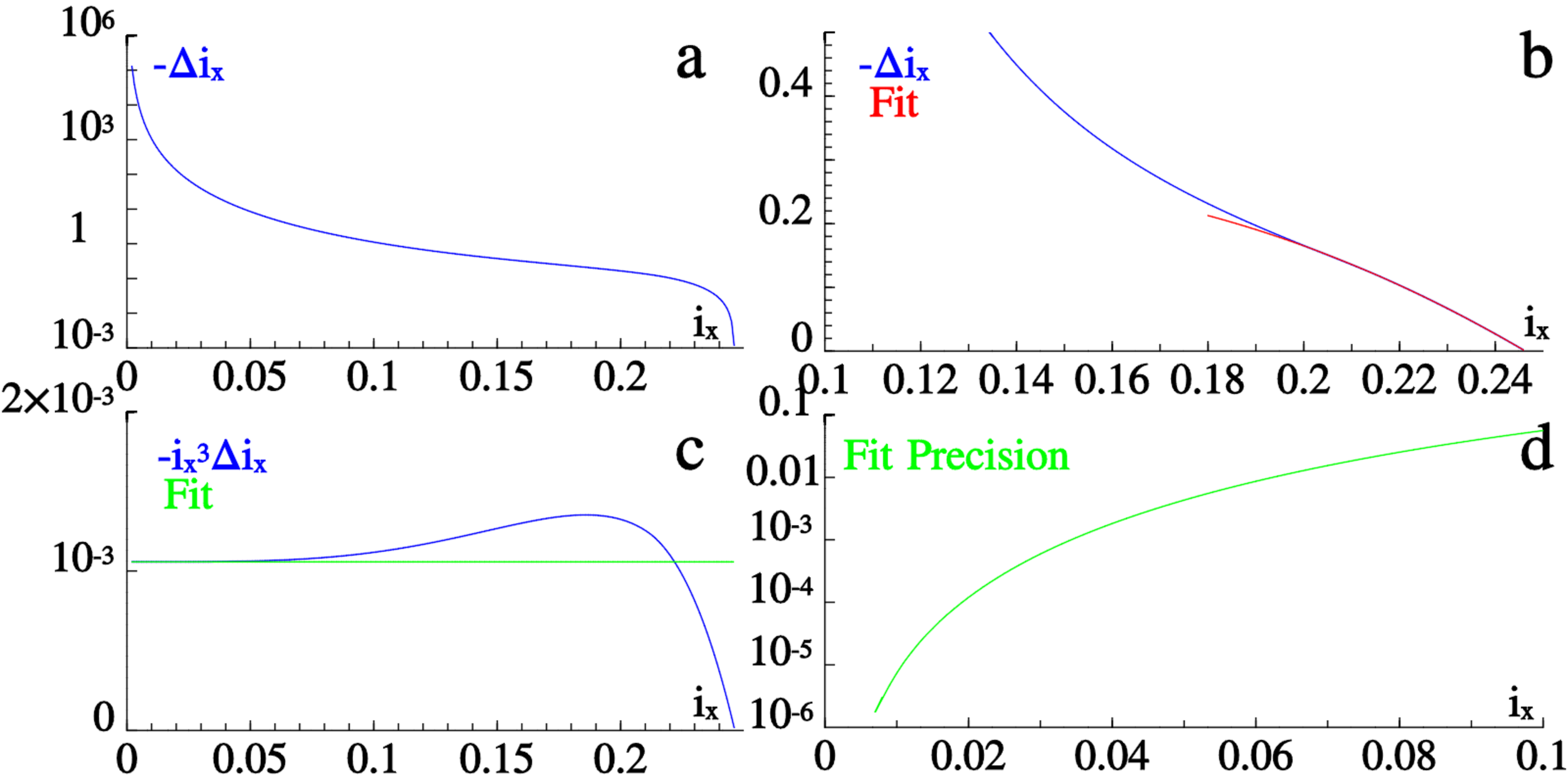}
\caption{\label{Fig:dix} (a) $-\Delta i_x(i_x)$ on a logarithmic scale. (b) The lower part of $-\Delta i_x(i_x)$ ($i_x>0.1$) on a linear scale (blue), its rough fit by a quadratic function (red). (c)  $-i_x^3\Delta i_x$ (blue) and its fit (green), in this case a constant. (d) Accuracy of the fit on a logarithmic scale.}
\end{figure*}

The graph of $-\Delta i_x(i_x)$ is shown in Fig.~(\ref{Fig:dix}). It is indeed evident that $i_x^3\Delta i_x=const$ is a good approximation for $i_x\ll i_{xmax}$. The rough fit near $i_x=i_{xmax}$ (in red) was obtained numerically and is $-\Delta i_x=4.4(i_{xmax}-i_x)-18(i_{xmax}-i_x)^2$.

\subsubsection{\label{sec:f2g2} $\Delta\mu(i_x)/\Delta i_x(i_x)$}

The function $\Delta\mu(i_x)/\Delta i_x(i_x)$ or $F_{i_x}(i_x)/G_{i_x}(i_x)$ is crucial for the calculation of the invariant $L$ according to Eq.\ref{eq:L1L2}. The graph depicts the negated value $-\Delta\mu(i_x)/\Delta i_x(i_x)$, or $-\Delta\mu/\Delta i_x$. $-\Delta\mu/\Delta i_x\approx2/i_x$ is expected near $i_x=0$.

\begin{figure*}
\includegraphics[width=2\columnwidth]{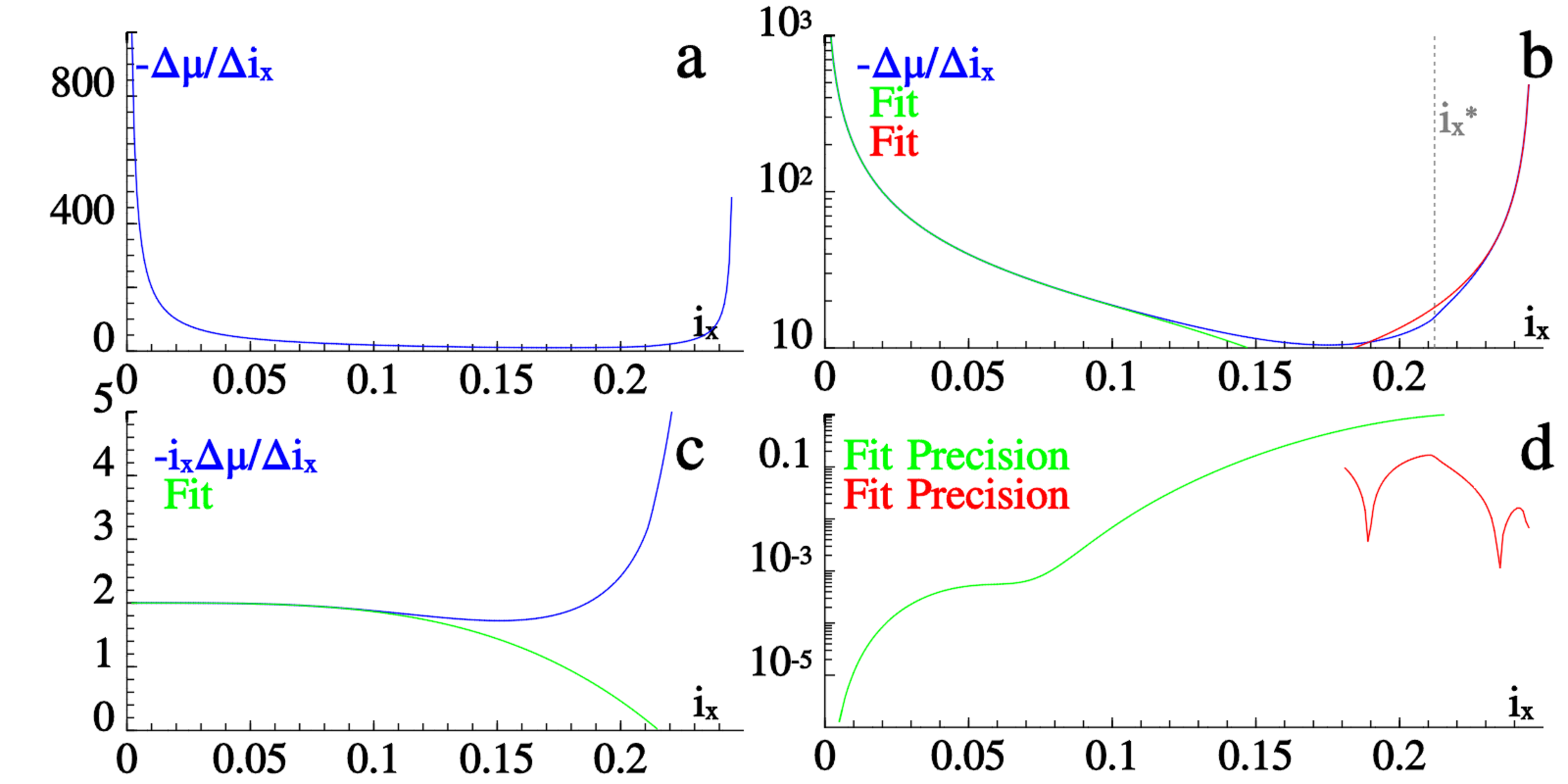}
\caption{\label{Fig:f2g2} (a) $-\Delta\mu(i_x)/\Delta i_x(i_x)$ on a linear scale. (b) $-\Delta\mu(i_x)/\Delta i_x(i_x)$ on a linear scale (blue), its fits at $i_x\ll i_{xmax}$ (green) and near $i_x=i_{xmax}$ (red), $i_x=i_x^*$ (gray). (c) $-i_x\Delta\mu(i_x)/\Delta i_x(i_x)$ (blue) and its fit at $i_x\ll i_{xmax}$ (green). (d) Accuracy of the fits on a logarithmic scale.}
\end{figure*}

The graph of $-\Delta\mu(i_x)/\Delta i_x(i_x)$ is shown in Fig.~\ref{Fig:f2g2}. The expected fit $2/i_x$ at $i_x\ll i_{xmax}$ works well, a more precise one obtained numerically and used in Fig.~\ref{Fig:f2g2} in green is $2/i_x-430i_x^{5/2}$. Near $i_x=i_{xmax}$ the linear character of $-\Delta i_x$ on Fig.\ref{Fig:dix} and the roughly constant character of $\Delta\mu$ on Fig.\ref{Fig:dmu} suggest a very rough fit can be made, shown in red and equal to $0.62/(i_{xmax}-i_x)$. On subplot (b) in the logarithmic scale a breaking point near $i_x=i_x^*$ is once again evident.

\subsubsection{\label{sec:L} $L(i_x)$}

The function $L(i_x)$ is obtained numerically using expression (\ref{eq:L1L2}). $L\approx i_x^2$ is expected at $i_x\ll i_x{max}$.

\begin{figure*}
\includegraphics[width=2\columnwidth]{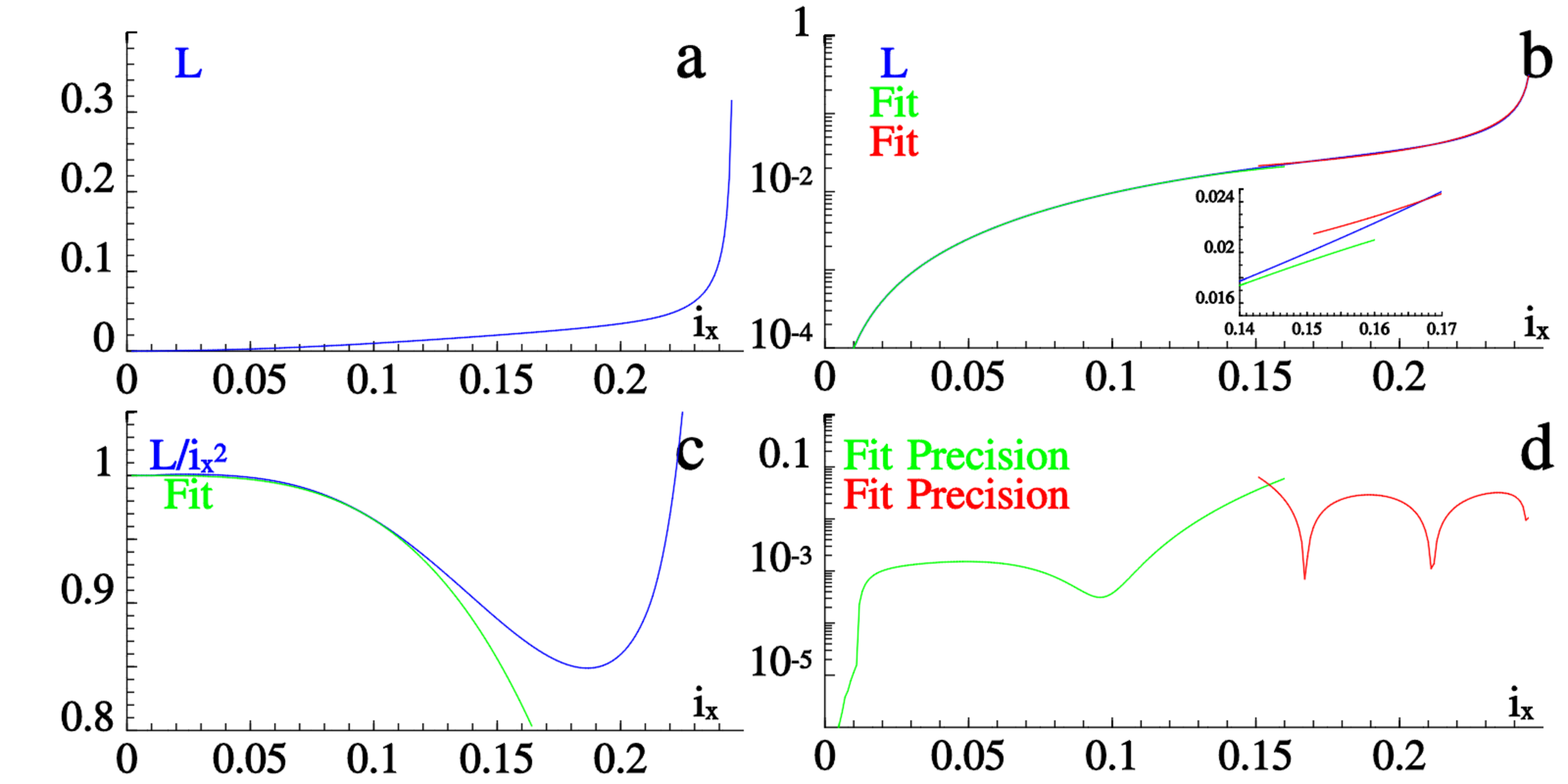}
\caption{\label{Fig:L} (a) $L(i_x)$ on a linear scale. (b) $L(i_x)$ on a logarithmic scale (blue), its fits at $i_x\ll i_{xmax}$ (green) and near $i_x=i_{xmax}$ (red). (c) $L(i_x)/i_x^2$ (blue) and its fit at $i_x\ll i_{xmax}$ (green). (d) Accuracy of the fits on a logarithmic scale.}
\end{figure*}

The graph of $L(i_x)$ is shown in Fig.~\ref{Fig:L}. The expected fit $L\approx i_x^2$ at $i_x\ll i_{xmax}$ works well, a more precise one used in Fig.~\ref{Fig:L} is $L=i_x^2-110i_x^{11/2}$. As follows from the fit to $\Delta\mu(i_x)/\Delta i_x(i_x)$, a rough fit can be made near $i_x=i_{xmax}$, shown in red and equal to $5\cdot10^{-3}(i_{xmax}-i_x)^{-0.62}$.

\subsubsection{\label{sec:takeaways} Takeaways}

In the assumptions of weak radiative friction (\ref{RRweak}) and ultrarelativity ($\gamma\gg1$ or $\mu\ll1$) the motion of particles in current sheets with the field structure (\ref{eq:B}) was studied. It is also assumed that the field component $B_x$ contributes only to the shape of the unperturbed trajectory, making it 3D instead of 2D \cite{Traj,ZelenyiUFN2013,Sonnerup1971}, but does not contribute to radiative friction. Since the former is a qualitative change and the latter is a quantitative contribution, such assumptions can be justified in the case when $B_x$ is sufficiently weak \cite{ZelenyiUFN2013}.

Within these assumptions the main parameters defining the trajectory ($\mu$ and $i_x$), being integrals of motion in the frictionless case, are now considered as slowly varying parameters. In order to quantify the influence of radiative friction, the change in these parameters due to radiative friction is integrated over a half-period $T_y/2$ of the unperturbed trajectory. The exact expressions even within our assumptions are rather intricate (See Eqs.(\ref{eq:dmu}),(\ref{eq:dix}),(\ref{eq:etaix}),(\ref{eq:f}),(\ref{eq:cos}),(\ref{eq:elliptic})), mostly owing to the implicit definition of the dual-branch function $f^{-1}$ and the piecewise properties of $f^{-1}$ and other underlying functions. Thus, the required functions have been nondimensionalized and calculated numerically.

The analytical approximations made in the limit $i_x<<i_{xmax}$ work well up to at least $i_x\approx0.05$, some up to $i_x\approx0.15$. When working in the specified limit, these may be sufficient depending on the required precision. At higher $i_x$ (or if precision better than $10^{-2}-10^{-3}$ is required), however, the numerically obtained functions can prove to be useful as the fits no longer work very well.

We would like to comment on the surprising change of trend near $i_x=i_x^*\approx0.212$ in $\Delta\mu(i_x)$ on Fig.\ref{Fig:dmu}(a,b) and the lack thereof on $-\Delta i_x(i_x)$ in Fig.\ref{Fig:dix}(a,b). The decrease of both with the growth of $i_x$ from 0 to $i_x^*$ is not surprising, since the part of the trajectory to the right of the \textit{uncertainty curve} is the most subject to radiation losses. We believe the slight increase in $\Delta\mu$ at $i_x>i_x^*$ is due to the fact that near the uncertainty curve the trajectory of the particle in the "fast" phase space is close to the separatrix, so the particle is less subject to radiation reaction \cite{Traj}, or, in mathematical terms, $\overline{x^2}\approx0$ in Eq.\ref{Frad2} or $\eta\approx\overline{\cos{\varphi}(\eta)}\approx-1$ in Eq.\ref{eq:dmu}. As a result, the increase in $\Delta\mu$ when moving away from the uncertainty curve is noticeable, but still rather slight. For $\Delta i_x(i_x)$, however, the factor $f'(-\eta)$ in (\ref{eq:dixdmu}-\ref{eq:dix}) mitigates this, since for trajectories close to the center of the "slow" phase space (Fig.\ref{phsp}, $i_x\approx i_{xmax}$) $f'(-\eta)\approx0$ as the phase space center corresponds to $\eta=\eta^*$ \cite{Traj}. See Supplemental Material at [URL will be inserted by publisher] for some useful approximations in the limit $i_x\approx i_{xmax}$.

Finally, we would like to expand on the influence of radiative friction on the two main parameters chracterizing the state of the particle. For the first parameter, $\mu=1/\gamma$, this influence is qualitatively obvious since friction is always dissipative: the particle's energy decreases, therefore $\gamma$ decreases and $\mu$ increases. For the parameter $i_x$, however, this influence is a bit more tricky. The value of the expression (\ref{eq:dixdmu}), as well as of the integrand of Eq.(\ref{eq:dix}), can be both positive or negative. This depends on the sign of $\eta f'(-\eta)$, and therefore is positive between the green curve and the vertical line $y=0$ on Fig.~\ref{phsp}, and negative otherwise. However, as it appears from Fig.~\ref{Fig:dix}, when integrated over a half-period of "slow" motion, the value of $\Delta i_x$ is strictly negative. This means that, if the condition (\ref{RRweak}) is satisfied, $i_x$ will decrease with the flow of time $t\gtrsim T_y$, the trajectories on the "slow" phase space Fig.~\ref{phsp} will spiral outwards from the center, and the trajectories in the "fast" $x-z$ plane will go from oscillating near the 8-shape to alternating between trajectories of type B and type C and then to oscillating between type A and D via the path A-B-C-D-C-B-A (see Refs.~\cite{Traj,ZelenyiUFN2013}).

\section{\label{sec:Disc} Discussion}

The method of slowly varying parameters was used to quantify the influence of radiative friction over a single half-period $T_y/2$. The obtained expressions and the performed numerical calculations can be used to evaluate the changes of key parameters $\mu$ and $i_x$ over $T_y/2$, to employ the obtained invariant $L(\mu,i_x)$ or to construct a system of differential equations describing the system on long timescales $\Delta t\gg T_y$. This system is much easier to analyze on the specified timescales than the original one (\ref{systembase}), because, first, it is a system of only two differential equations instead of five or six, and second, one would need to resolve only the time scales $\mu/\dot{\mu}$ and $i_x/\dot{i_x}$ instead of $T_x$, which is at least two orders of magnitude lower (\ref{RRweak}).

The primary condition for the applicability of the entire model described within this paper is $T_x/T_y\ll1$. Using this model $T_y$ can be expressed as $T_y=\int{dq_y/\dot{q_y}}=T_y(i_x)$, and using the model from \cite{Traj} $T_x=\int{d\varphi/\dot{\varphi}}=T_x(\eta(i_x,q_y))$ can be written. In other words, $T_y$ depends on which trajectory of the phase plane (Fig.2) the particle is located on, while $T_x$ also depends on which point of the trajectory it is currently at. Thus, the constructed theory allows us to obtain $T_x/T_y$ as a function of $(i_x,q_y)$ or $(y,q_y)$. In the region to the left (and far enough from) the uncertainty curve, it was found that the condition can be reduced to $(kmc^2)/(\mu eB_x^2)\gg1$, which is similar to the condition obtained in \cite{ZelenyiUFN2013}. To the right of the uncertainty curve $T_x\sim1/\sqrt{-\eta}$, i.e. is much less, so if $T_x/T_y\ll1$ is satisfied on the part of trajectory to the left from the uncertainty curve, it will be automatically satisfied to the right of the uncertainty curve as well.

Additionally, while we used the method of slowly varying parameters to calculate the according functions in the assumptions described in Sec.\ref{sec:class}, it can be applied to a more general problem, where some of these assumptions can be completely or partially lifted.

The second and third terms in (\ref{dotmu}) involving $B_x$ do not have to be neglected. Finding the exact expressions for $\Delta\mu$ and $\Delta i_x$ would entail somewhat more intricate math, but the method applies. Interestingly, the form of the invariant $L$ in this case remains the same. Note that the condition for weakness of $B_x$\cite{ZelenyiUFN2013} is still required in order to be able to describe trajectories with the model described in Sec.\ref{sec:base} in the first place.

Finally, with some modifications to the method the requirement $T_y\ll{}\gamma/\dot{\gamma},i_x/\dot{i_x}$ (\ref{RRweak}) can be lifted (but not $T_x\ll{}\gamma/\dot{\gamma},i_x/\dot{i_x}$ or $T_x\ll T_y$). In this case differential equations for $\mu$ and $i_x$ have to be solved self-consistently alongside with the one for $q_y$ and would describe motion on time intervals within the period $T_y$.

We note that although technically the presented method can be used with some of these assumptions lifted simultaneously, it would require immense analytical and computational efforts. Therefore, we remind that one should adapt to the problem at hand and lift only those assumptions that truly require doing so.

Lastly, the effect of quasirandom jumps of the quasiadiabatic invariant of nondissipative motion $i_x$ \cite{ZelenyiUFN2013} in the vicinity of the uncertainty curve were ignored in this work. Since they occur in a close vicinity of the uncertainty curve during time intervals $\Delta t\ll T_y$, the change to $\mu$ during this time can be neglected. Therefore, the quasiadiabatic invariant of dissipative motion $L(\mu,i_x)$ will experience jumps according to the jumps in $i_x$, which have been extensively studied before.

\section{\label{sec:conc}Conclusion}

The motion of an ultrarelativistic particle experiencing radiative friction in null current sheets with a two-component magnetic field (with a weak component perpendicular to the sheet) was considered. The method of slowly varying parameters was used to quantify the influence of weak radiation reaction considered in the form of the continuous Landau-Lifshitz force on key parameters that define the dynamics of the particle. The expressions quantifying this influence were found, and due to their intricacy they were nondimensionalized and computed numerically. A quasiadiabatic invariant of dissipative motion was found analytically and also computed numerically in dimensionless variables. Theoretical fits were tested, their precision analyzed depending on the value of key parameters, and in some cases more precise fits were found numerically.

It was also shown how the presented method can be applied in a wider range of problems where some of the employed assumptions are completely or partially lifted. 

\begin{acknowledgments}
The work was supported by the Ministry of Science and High Education of the Russian Federation under agreement No.~075-15-2021-1361.
A.A.M. acknowledges the support of the Foundation for the Advancement of Theoretical Physics and Mathematics "BASIS".
\end{acknowledgments}

\bibliography{bibliography}

\end{document}


	
	\title{Dynamics of radiating particles in current sheets with a transverse magnetic field component:\\Supplemental Material}
	
	
	\author{A.~A.~Muraviev}
	\email{sashamur@gmail.com}
	\affiliation{Institute of Applied Physics, Russian Academy of Sciences, Nizhny Novgorod 603950, Russia}
	
	
	

	
	\date{\today}
	\maketitle

    \section{\label{Appendix} Approximations in the $i_x\approx i_{xmax}$ limit}

The limit $i_x \rightarrow i_{xmax}$ corresponds to the neighborhood of the center of the phase plane $(y,q_y)$, and the structure of the phase plane (Fig.2 \footnote{Here and below references to figures or equations refer to ones from the main text}) suggests that near this point it is possible to linearize the system of differential equations (4) to $\dot{y}$ and $\dot{q_y}$. Before doing this, it is necessary to take a closer look at the function $\overline{x^2}(q_y,\overline{\cos{\varphi}}(\eta(y,q_y)),\eta(y,q_y ))$ and the functions contained therein that need to be linearized, and their values at the point corresponding to $i_x=i_{xmax}$. Since the curve $\dot{q_y}=0$ passes through this point (see Fig.1), then, given (15), $\overline{\cos{\varphi}}=0$. From the latter it follows that $\eta=\eta^*\approx-0.65$ \cite{Traj}, and $\overline{x^2}$ takes some non-zero value. Using the expressions (4-8,15) and leaving only the terms up to the first order of magnitude by $y-y^*$ and $q_y$, the system can be linearized and any necessary parameters of the system oscillations near the center can be obtained. Among other things, the frequency of these oscillations $\omega_0\approx\sqrt{0.87}eB_x\mu/mc$ was obtained, which is of interest for further research, where the coefficient is equal to the derivative of $\overline{\cos{\varphi}}$ with respect to $\eta$: $\cos{\varphi}'_\eta\approx0.87$ which appears during linearization. Thus, the half-period is $\pi/\sqrt{\cos{\varphi}}'_\eta\approx3.37$ in dimensionless units, which agrees well with the values obtained numerically near $i_x=i_{xmax}$.

An interesting method of calculating approximations can be used in the $i_x \rightarrow i_{xmax}$ limit. For this method we switch back to integrating over $dt$ as opposed to $dq_y$ as proposed withing this paper. As we will see later, many of the functions of interest to us are functions of $\eta$ and $q_y^2$. Since $\overline{\cos{\varphi}}\rightarrow0$ and $q_y\rightarrow0$ in this limit, all appearances of $\eta$ can be expressed via the Taylor series in terms of $\overline{\cos{\varphi}}\equiv w$. Then $\eta=\eta^*+\eta'_ww+\eta''_{ww}w^2/2+...$. Here higher terms are neglected. As follows from Eq.15, any integral over a whole or half-period involving the first power of $\overline{\cos{\varphi}}$ is equal to 0. As a result, we are left with terms of the order 0 (constant) and 2 (quadratic). The quadratic term can then be expressed back through $\eta$ and replaced with a Taylor expansion of Eq.16 over $\eta$. As a result, the remaining terms will be either constant or quadratic over $q_y$ with no other variable that changes within a $i_x=const$ trajectory. For quadratic terms it is well known that during harmonic oscillations their average value is half of the max value, which can be easily expressed linearly over $(i_{xmax}-i_x)$ using Eq.9. This method allows calculating many integrals in this limit without actually performing integration.

\bibliography{bibliographySupp}